\documentclass[10pt]{iopart}

\usepackage{iopams}  
\usepackage{cite} 
\usepackage{hyperref} 
\hypersetup{colorlinks=true, urlcolor=blue, linkcolor=blue, 
citecolor=blue} 
\usepackage{graphics} 
\usepackage{graphicx} 
\usepackage{amssymb}

\begin{document}

\title[Magnetic properties of an antiferromagnetic spin-1/2 XYZ model ]{Magnetic properties of an antiferromagnetic spin-1/2 XYZ model in the presence of different magnetic fields: finite-size effects of inhomogeneity property}

\author{Hamid Arian Zad}

\address{Young Researchers and Elite Club, Mashhad Branch, Islamic Azad University, Mashhad, Iran}%
\address{Alikhanyan National Science Laboratory, Alikhanian Br. 2, 0036 Yerevan, Armenia}%
\eads{\mailto{\normalfont \color{blue} arianzad.hamid@mshdiau.ac.ir}}

\author{Azam Zoshki}%
\address{Young Researchers and Elite Club, Mashhad Branch, Islamic Azad University, Mashhad, Iran}%
\eads{\mailto{\normalfont \color{blue} zoshkia@yahoo.com}}

\author{Moones Sabeti}%
\address{Young Researchers and Elite Club, Mashhad Branch, Islamic Azad University, Mashhad, Iran}%
%\eads{\mailto{\normalfont \color{blue} moones.sabeti@gmail.com}}

\vspace{10pt}
%\begin{indented}
%\item[]March 2015
%\end{indented}

\begin{abstract}
 Magnetic and thermodynamic properties of the anisotropic XYZ spin-1/2 finite chain under both homogeneous and inhomogeneous magnetic fields are theoretically studied at low temperature. Using exact diagonalization method (ED), we study the magnetization, magnetic susceptibility and specific heat of the model  characterized in terms of the finite correlation length in the presence of three different magnetic fields including longitudinal, transverse and transverse staggered magnetic fields.
 The magnetization, susceptibility and the specific heat of the model are investigated under two conditions separately: I) when the model is putted in the presence of homogeneous magnetic fields; II) when finite inhomogeneities are considered for all applied magnetic fields in the Hamiltonian.  
 We show that for the finite-size XYZ chains at low temperature, the evident magnetization plateaus gradually convert to their counterpart quasi-plateaus when the transverse magnetic field increases.
 Moreover, the influence of the transverse and staggered transverse magnetic fields, and their corresponding inhomogeneities on the magnetization process, magnetic susceptibility, and specific heat are reported in detail. Our exact results illustrate that by altering the inhomogeneity parameters, magnetization plateaus gradually convert to their counterpart quasi-plateaus. The specific heat manifests Schottky-type maximum, double-peak and triple-peak, as well as, transformation between them by varying considered inhomogeneity parameters in the Hamiltonian.

\end{abstract}
%\submitto{\jpa}
%\maketitle
\section{Introduction} \label{sec:level1}
Exactly solved one-dimensional (1-D) spin models represent important milestones in statistical mechanics, since they pave the way to understand several aspects of magnetic materials in the real world. The spin S = 1/2 Heisenberg models (XX, XYZ, XXZ) in the presence of a longitudinal magnetic field are paradigmatic examples of exactly tractable models,  which not only have been applicable to elucidate generic features of quantum phase transitions, but also have long served as a paradigm for the study of quantum magnetism in low dimensions \cite{Thakur2018,Mikeska2004}. The study of external homogeneous magnetic field influences on the Heisenberg spin-1/2 models have been encountered with a lot of attentions in terms of both theoretical and experimental condensed matter physics \cite{Zad2018,Zad2017,str15,Dmitriev12002,Dmitriev2004,Dmitriev22002,Hagemans2005,Hikihara2004,Ovchinnikov2003,Caux2003,Enderle2007}.

The nonuniform magnetic field is rarely taken into account. It is  obvious that in any condensed matter physics subject, inhomogeneous zeeman coupling has remarkable effects on the energy band gaps as well as thermodynamic parameters of the quantum spin systems. So it is momentous to investigate the thermodynamic behavior of a spin system under a nonuniform magnetic field. Recently, M. Panti{\'c} {\it et al.} \cite{Pantic2017} studied the effect of inhomogeneous  magnetic field on the thermodynamic properties of an isotropic two-qubit XXX spin system. We note that the magnetization behavior for a XXZ spin model in nonuniform magnetic fields either longitudinal or transvesal has not been discussed specifically at low temperature. Although Felicien Capraro and Claudius Gros \cite{Capraro2002} studied the influence of both homogeneous longitudinal and transverse fields as well as transverse staggered field on opening of a spin-gap in 1-D spin chain, in the theoretical analysis we are strongly believe that it is stimulating and should be investigated  the magnetic and thermodynamic properties of the spin chain under  inhomogeneous magnetic fields specifically  inhomogeneous transverse staggered field. This is the main motivation of this paper.

To investigate the critical points \cite{Mikeska2004,Dmitriev12002,Giamarchi1988,Sato2004} in which phase transition occurs, the magnetic and thermodynamic properties of various metal containing compounds have been studied in the literature. Some of these compounds are very similar to 1-D spin-1/2  models. For instance, S. Eggert investigated magnetic and thermodynamic properties of material $Sr_2CuO_3$ in Refs. \cite{Eggert1994,Eggert1996}. It was demonstrated that both materials $Sr_2CuO_3$ and $SrCuO_2$ can be regarded as 1-D S-1/2 Heisenberg systems by fitting the temperature dependence of magnetic susceptibility with the theoretical calculation by Eggert, Affleck, and Takahashi (EAT) at low temperatures as low as $0.01J$ \cite{Motoyama1996}. 
The magnetic properties of rare-earth compound $Yb_4As_3$ in the absence of external fields, can be investigated by considering such compound as an antiferromagnetic Heisenberg spin-1/2 chain. The 4f-compound $Yb_4As_3$ in the vicinity of external homogeneous, longitudinal, transverse, and transverse staggered  magnetic fields  have profoundly been studied \cite{Affleck1997,Oshikawa1999,Capraro2002}. 

The specific heat behavior with respect to the temperature of spin S=1/2 chains has been studied by several groups \cite{Kikuchi2005,Rule2008,Matysiak2013,Abouie2006,Cond,NanSi2018,KaileShi2018}. In Ref. \cite{Johnston2000}, D. C. Johnston {\it et al.} indicated that parameter fluctuation effects play an essential role in the specific heat behavior versus temperature for  an insulator $NaV_2O_5$, which its magnetic susceptibility is that of a 1-D Heisenberg chain \cite{Gros1999}. They demonstrated a good agreement between theoretical results and experimental data. Furthermore, O. Breunig \cite{Breunig2013} experimentally studied the specific heat of one-dimensional magnetic material $Cs_2CoCl_4$ with a comparison to the theoretical predictions of the Heisenberg spin chain. Generally, they found a good description of the experimental analysis in high temperature and strong magnetic field, although some differences between theory and experiment were observed at finite magnetic field. 
The magnetic part of the specific heat can be usually estimated in a certain temperature range by the Schottky theory. The associated round maximum of the specific heat, the so-called Schottky peak (maximum), has been experimentally detected in various magnetic compounds
among which one could mention the molecular magnets. Recently, it has been theoretically reported that spin clusters can also depict the mentioned Schottky peak due to a typical competition between antiferromagnetic interactions and magnetic field \cite{Hucht2011,str16}.

In order to figure out the magnetic behavior of the spin-1/2 Heisenberg chains in terms of applied magnetic field and/or exchange couplings between spins, magnetization plateaus have considerably been examined for various copper oxide compounds \cite{Eggert1994,Motoyama1996,Oshikawa1997,Leiner2018,NanSi2018,KaileShi2018,Xing2019,Zhao2019}. The behavior of uniform magnetization in the different phases with their dependence on the longitudinal (transverse) field for fixed values of other applied parameters was studied by P. Thakur {\it et al.} \cite{Thakur2018}. Consequently, they observed that in the presence of the transverse field, the nature of the behavior of the uniform and staggered magnetizations near the critical fields dramatically change.
K. Hida obtained the magnetization curve by numerical diagonalization of finite size systems. The result explains the low temperature magnetization data for $3CuCl_2\cdot 2dx$. It is predicted that the magnetization curve has a plateau at 1/3 of the saturation magnetization if the ferromagnetic exchange energy is comparable to or smaller than the antiferromagnetic exchange energy \cite{Hida1994,Ajiro1994}. The magnetization curve as well as magnetic susceptibility has been measured by numerical diagonalization of finite size systems for material $3CuCl_2\cdot 2dx$. 
%An explicit magnetization plateau at $M_z=1/3$ of the saturation magnetization $M_s$ was observed for this material \cite{Hida1994,Ajiro1994}.

In this work, we will study the magnetic and  thermodynamic properties of a 1-D spin-1/2 chain in the presence of various kinds of applied homogeneous magnetic fields such as  longitudinal, transverse, and transverse staggered fields at low temperature. Then, we consider a finite inhomogeneity property for all applied magnetic fields and repeat our investigations, and compare our results with the case when the system is in the presence of homogeneous magnetic fields. We will limit our particular attention to a detailed examination of the magnetization, magnetic susceptibility and the specific heat. The plan of our paper is as follows: In the next section, we briefly discuss the XXZ model in the presence of the desired magnetic fields, and introduce the model by means of a well-understanding Hamiltonian. In section \ref{TM}, we discuss the behavior of thermodynamic parameters such as magnetization, magnetic susceptibility and specific heat and their dependences on the either homogeneous or inhomogeneous external fields. Finally, we end in section \ref{conclusions} with a brief summary and outlook.
\section{Model}\label{Model}
%The Heisenberg Hamiltonian is the only $SU(2)$-invariant nearest-neighbor spin Hamiltonian for spin-1/2 particles. 
\begin{figure}
\begin{center}
\resizebox{0.3\textwidth}{!}{%
\includegraphics{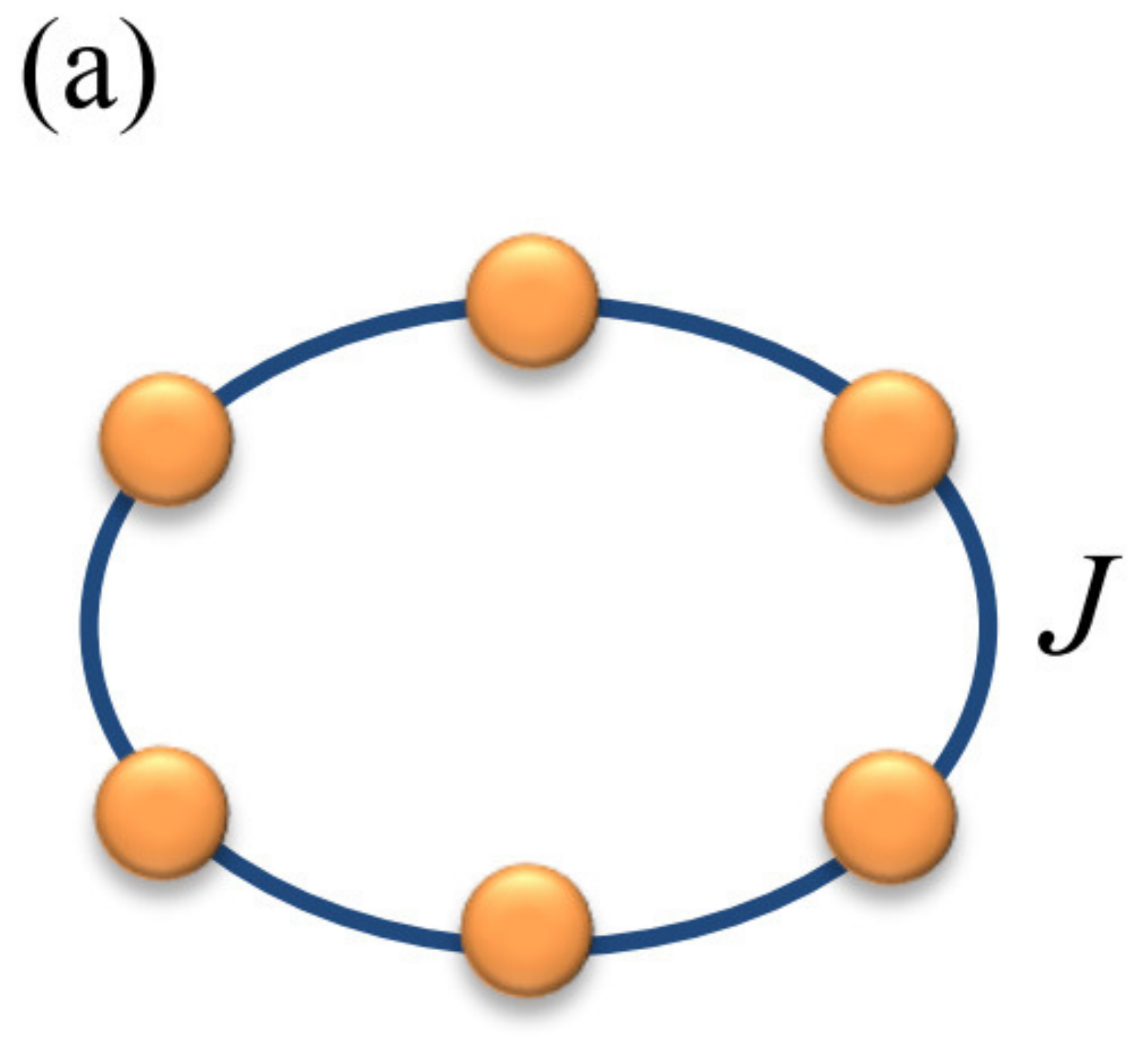}
}
\resizebox{0.45\textwidth}{!}{%
\includegraphics{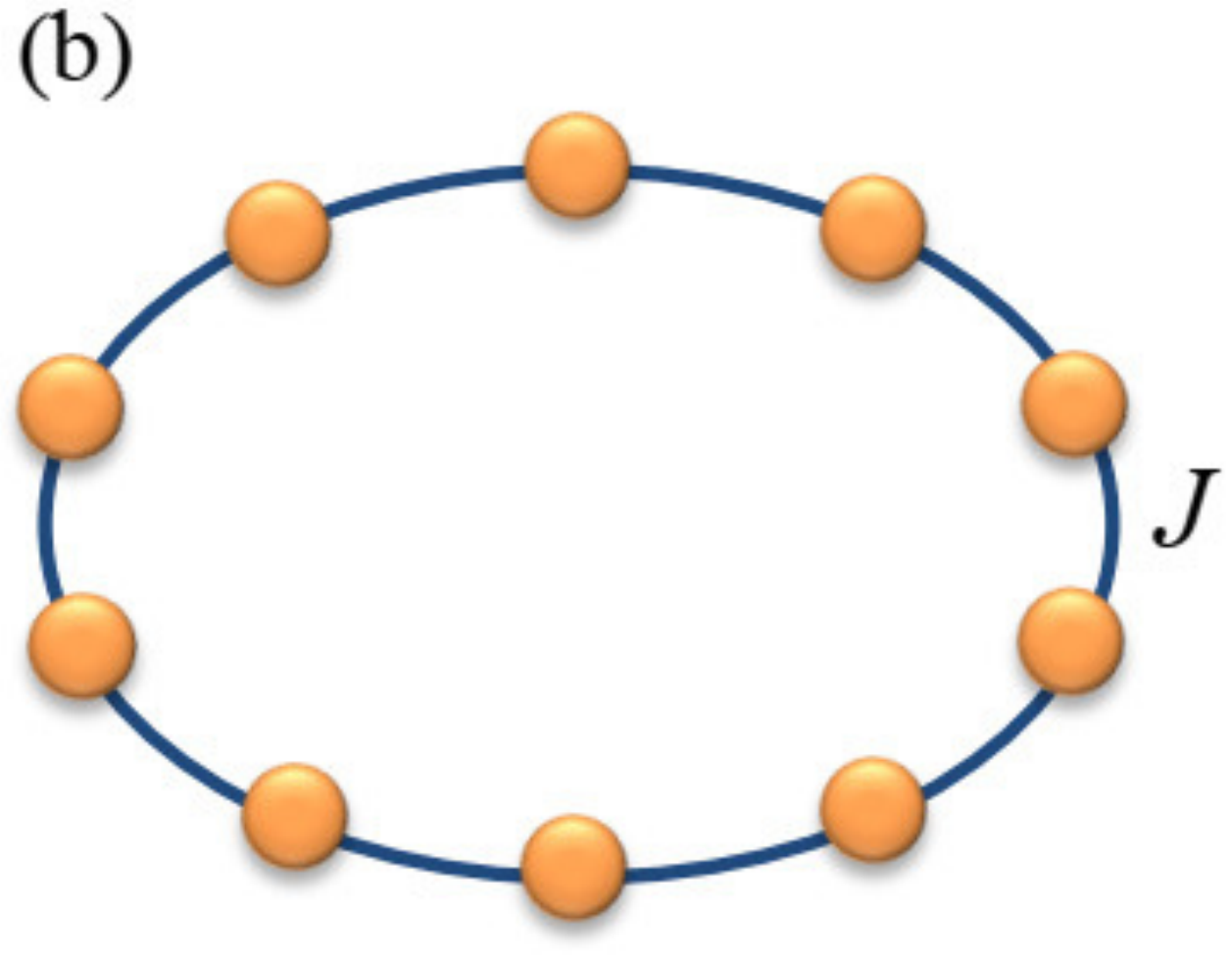}
}
\caption{Schematic representations of the spin-1/2 XYZ chain with finite length of (a) 6 particle, and (b) 10 particle, under periodic boundary conditions. $J$ denotes XYZ Heisenberg exchange interaction between each adjacent spins.}

\label{fig:SpinModel}
\end{center}
\end{figure}
The anisotropic S = 1/2 XYZ finite Heisenberg chain (Fig. \ref{fig:SpinModel}) as an exactly solvable system under inhomogeneous longitudinal and transverse, further transverse staggered magnetic fields, can be described by Hamiltonian
\begin{equation}\label{hamiltonian}
\begin{array}{lcl}
H_{XYZ}=\sum\limits_{j=1}^N{J}_{x} {S}_{j}^x{S}_{j+1}^x+ {J}_{y} {S}_{j}^y{S}_{j+1}^y+{J}_z {S}_{j}^z{S}_{j+1}^z\\
-\big[\sum\limits_{j=odd}(B_z+b_z){S}_{j}^z+\sum\limits_{j=even}(B_z-b_z){S}_{j}^z\big]\\
-\big[\sum\limits_{j=odd}(B_x+b_x){S}_{j}^x+\sum\limits_{j=even}(B_x-b_x){S}_{j}^x\big] \\
+\big[\sum\limits_{j=odd}(B_x^{stag}+\lambda)(-1)^j{S}_{j}^x+\sum\limits_{j=even}(B_x^{stag}-\lambda)(-1)^j{S}_{j}^x\big].
\end{array}
\end{equation}
%\begin{array}{lcl}
%H_{XXZ}=\sum\limits_{i=1}^N{J}_{xy} ({S}_{i}^x{S}_{i+1}^x+ {S}_{i}^y{S}_{i+1}^y)+{J}_z {S}_{i}^z{S}_{i+1}^z\\
%-\sum\limits_{i=1}^N\big[(B_z+b_z){S}_{i}^z+(B_z-b_z){S}_{i+1}^z\big]
%-\sum\limits_{i=1}^N\big[(B_x+b_x){S}_{i}^x+(B_x-b_x){S}_{i+1}^x\big] \\
%+\sum\limits_{i=1}^N\big[(B_x^{stag}+\lambda)(-1)^i{S}_{i}^x+(B_x^{stag}-\lambda)(-1)^i{S}_{i+1}^x\big].
%\end{array}
The integers $ j={(1,2,3,...,N)} $ are the number of spins, where under periodic boundary conditions: $ N+1=1$.  $J_{x}$, $J_{y}$ and $J_z$ represent the Heisenberg exchange interactions between adjacent spins $S_j$ and $S_{j+1}$ ($S^{\alpha}$ with $\alpha=\{x, y, z\}$ are spin-1/2 operators), and the sum is over unique exchange bonds.
$B_z$ is uniform longitudinal magnetic field, $B_x$ represent transverse field, and $B_{x}^{stag}$ denotes staggered transverse field incorporates all features proposed to be relevant for real materials like $Yb_4As_3$. The applied magnetic fields include the gyromagnetic g-factors and Bohr magneton  coefficient. parameters $b_z$, $b_x$ and $\lambda$ control the degree of inhomogeneity imposed into the longitudinal, transverse, and transverse staggered  fields, respectively. We note that according to our assumption, the inhomogeneity  leads to difference in strength of the induced magnetic fields into the odd and even sites of the Hamiltonian.

The transverse staggered  magnetic field applied in the Hamiltonian is directly induced by a staggered Dzyaloshinsky-Moriya (DM) interaction given by  \cite{Affleck1997}
\begin{equation}\label{DM}
\begin{array}{lcl}
 \sum_{j}(-1)^j{\bf D}\cdot({\bf S}_{j}\times {\bf S}_{j+1}), 
 \end{array}
\end{equation}
in which ${\bf D}$ is the length of the DM vector (here we consider the $z$-direction). Supposing  $D = |{\bf D}| = J_z sin(2\theta)$ the DM interaction can be eliminated by a rotation around ${\bf D}$ by an angle $\theta$ leading to  $B^{stag}_x = sin(\theta)B_z$, which can be interpreted as an effective staggered g-tensor.

It is quite obvious that the effect of a homogeneous longitudinal field like $B_N^h=-B_z\sum_{j=1}^N{S}_{j}^z$ on the structure of the XYZ spin chain, is not too much. This can be easily understood by noticing that $[H_{XX},B_N^h]=0$, where $H_{XX}$ is the Hamiltonian of an XX spin chain in the absence of external magnetic field \cite{Genovese2011}. What is really fascinating is applying an inhomogeneous longitudinal field defined by 
\begin{equation}\label{StaggMF}
\begin{array}{lcl}
B_N^I=-\sum\limits_{j=1}^NB_z(j){S}_{j}^z,
\end{array}
\end{equation}
for which generic magnetic field $B_z(j)$ is dependent on the site $j$. In this case Eq. (\ref{StaggMF}) does not commute with the total Hamiltonian of the system, namely
\begin{equation}\label{Commut}
\begin{array}{lcl}
[H_{XY},B_N^I]=2i\sum\limits_{j=1}^N\big[B_z(j+1)-B_z(j)\big]({S}_{j}^y{S}_{j+1}^x-{S}_{j}^x{S}_{j+1}^y).
\end{array}
\end{equation}
By performing some straightforward calculations, one can prove that the sum of all inhomogeneous magnetic fields applied in Eq. (\ref{hamiltonian}) does not commute with the total Hamiltonian. Consequently, the important feature of the Hamiltonian (\ref{hamiltonian}) is its the noncommutativity with the magnetization operator. This non-commutativity leads to a non-linear transverse magnetic field dependence of the spectrum of the model and to the phenomena of quasi-plateau in magnetization curve \cite{Ohanyan2018}. Regarding this, we here assume that the system under consideration is in the presence of external inhomogeneous magnetic fields as specified in Eq. (\ref{hamiltonian}).
\section{Results}\label{TM}
\begin{figure}
\begin{center}
\resizebox{0.4\textwidth}{!}{%
\includegraphics{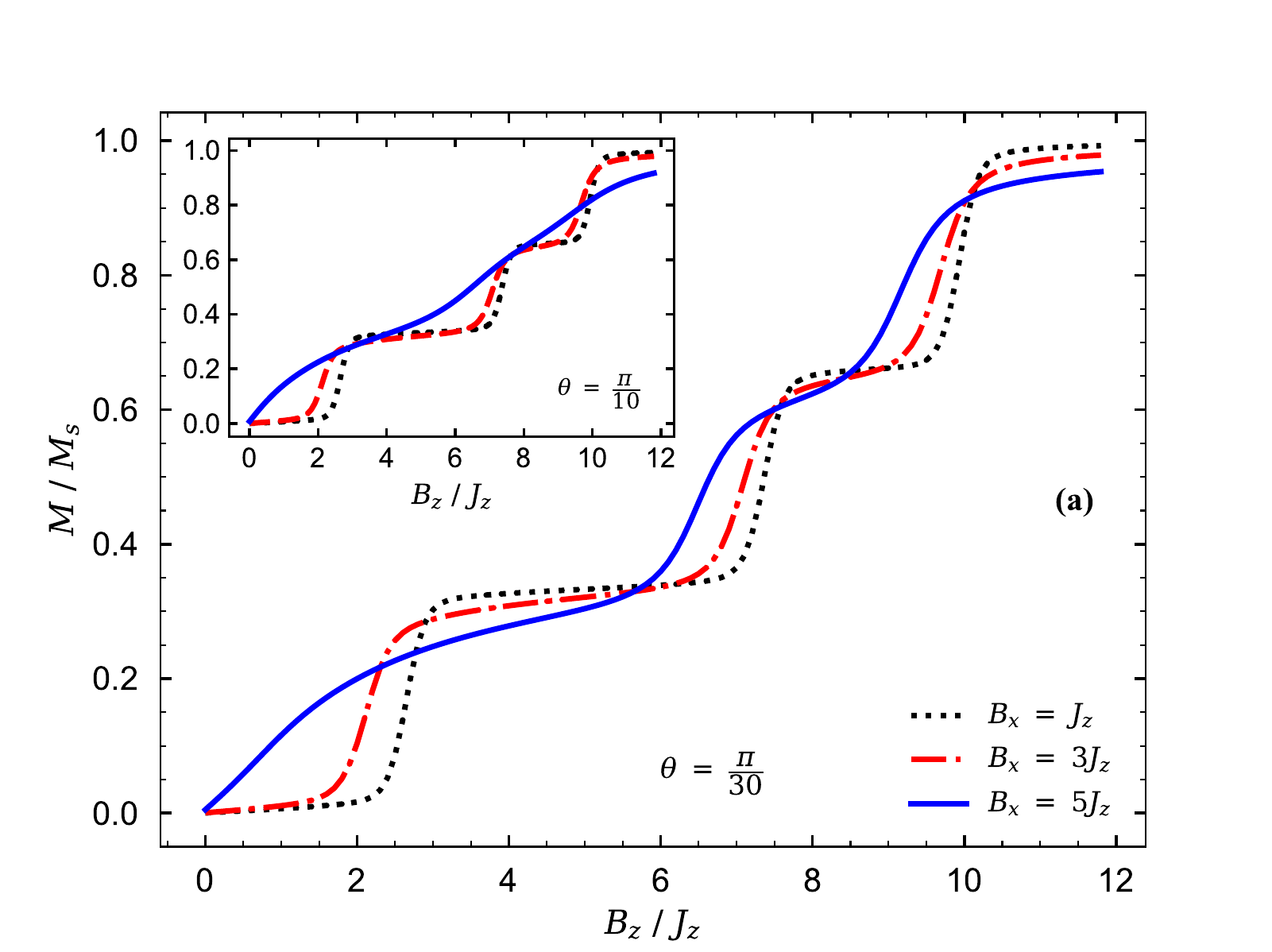}
}
\resizebox{0.4\textwidth}{!}{%
\includegraphics{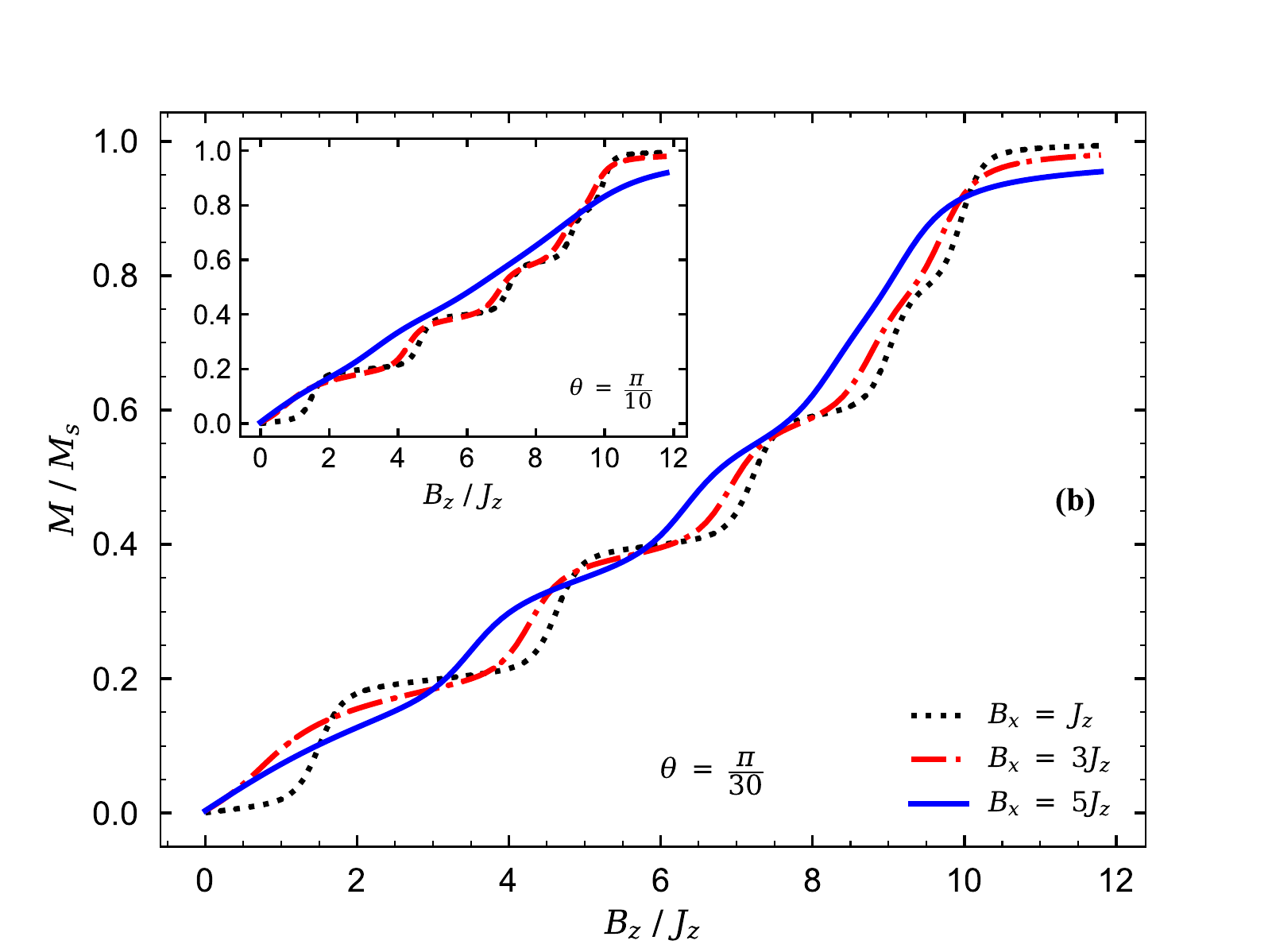}
}
\resizebox{0.4\textwidth}{!}{%
\includegraphics{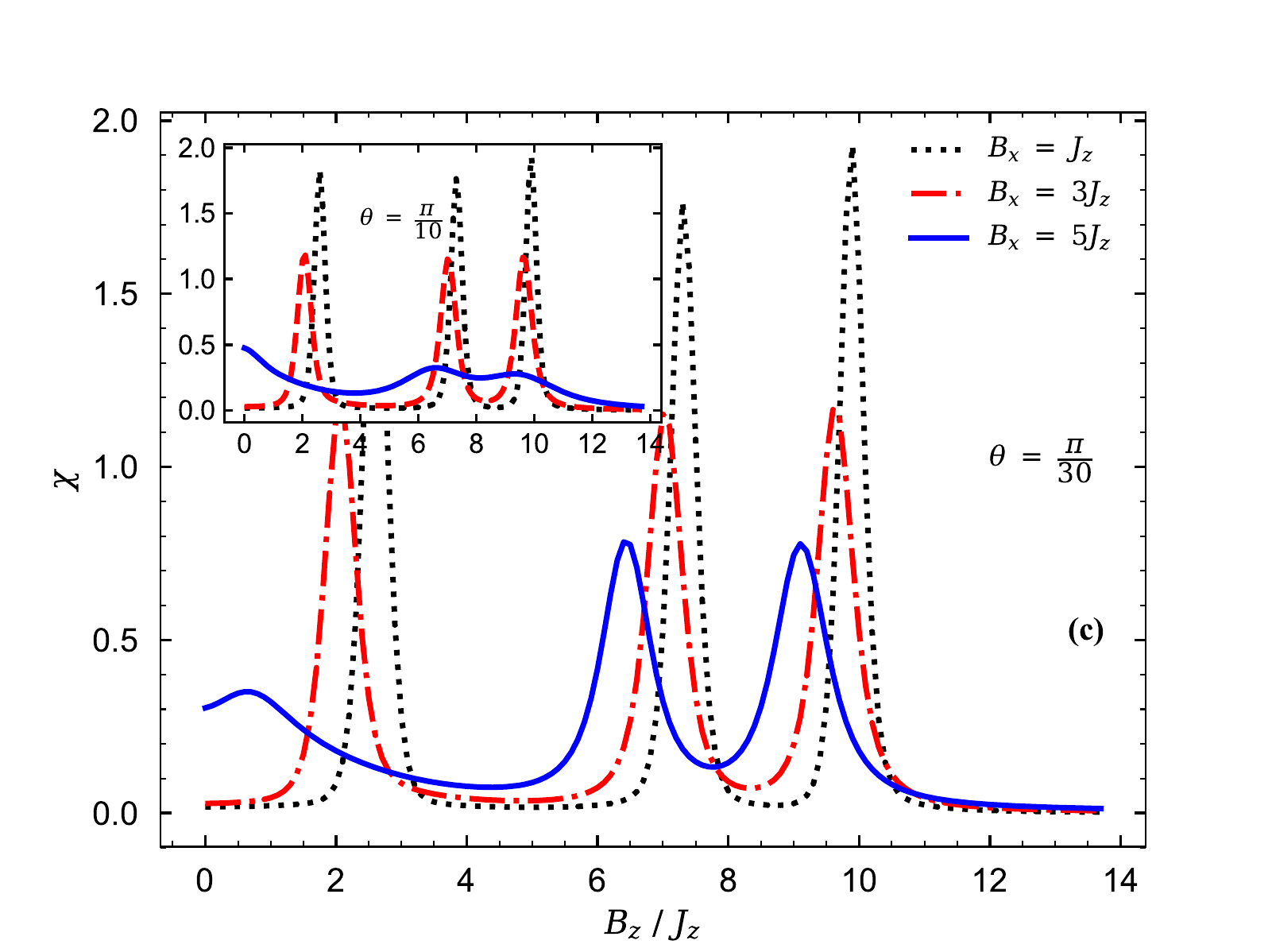}
}
\resizebox{0.4\textwidth}{!}{%
\includegraphics{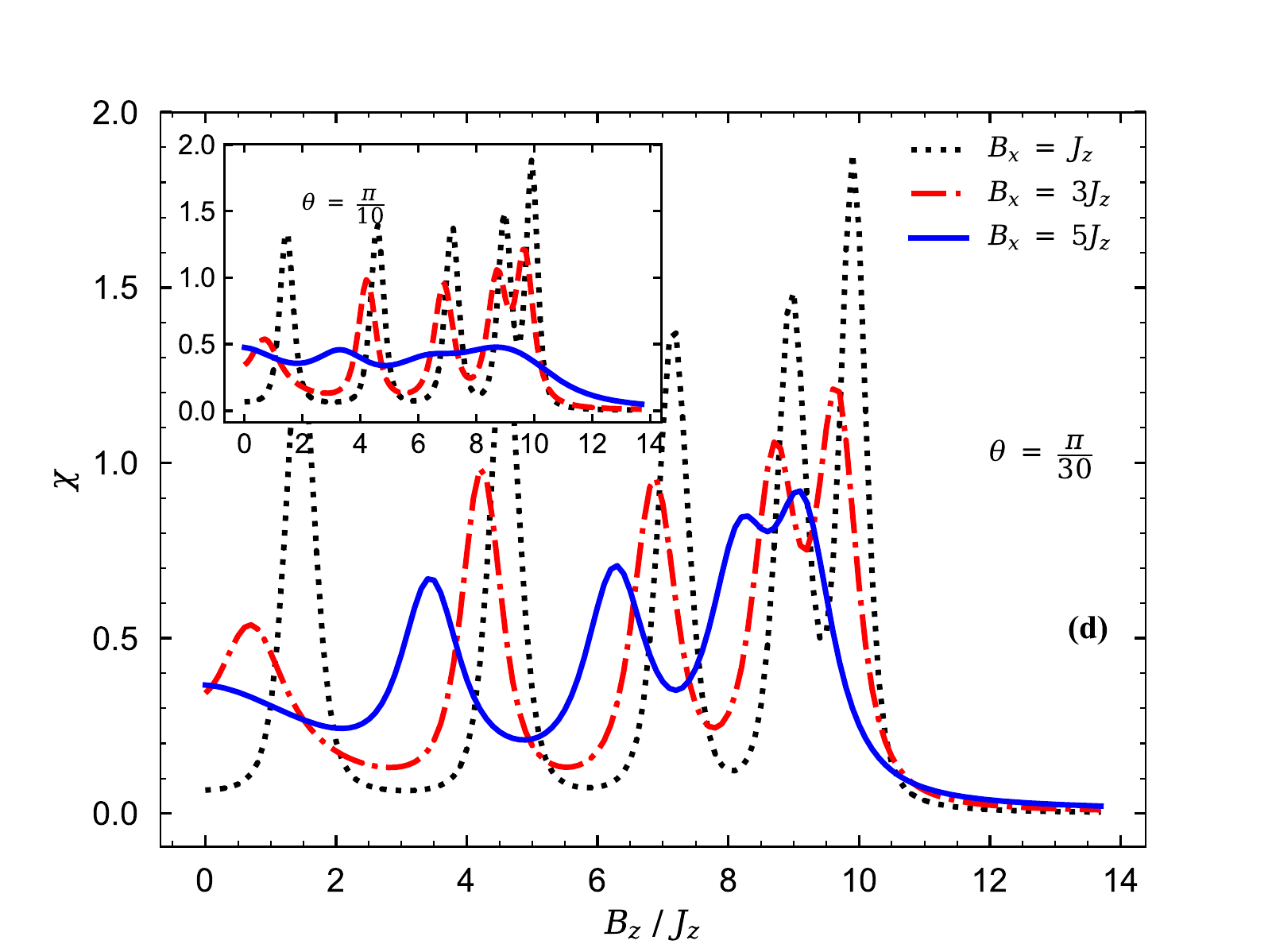}
}
\caption{Low-temperature $(T=0.1J_z)$ magnetization and magnetic susceptibility as functions of the longitudinal magnetic field $B_z/J_z$ for several fixed values of the transverse field $B_x/J_z$ , and an arbitrary angle $\theta=\frac{\pi}{30}$ under the condition $b_z=b_x=\lambda=0$. The system is considered in the presence of external homogeneous magnetic fields for which coupling constants have been conventionally taken as $J_{x}=8J_z$ and $J_y=10J_z$. Panels (a) and (c) correspond to the spin-1/2 XYZ model of finite length $N=6$; panels (b) and (d) correspond to the chain of length $N=10$. Insets show the corresponding magnetization and magnetic susceptibility curves for different angle $\theta=\frac{\pi}{10}$.}
\label{fig:MXB1}
\end{center}
\end{figure}

 In the present work, firstly, we examine in detail magnetic fields dependences of the magnetization, magnetic susceptibility and specific heat of the model introduced by Eq. (\ref{hamiltonian}) with the uniform exchange interactions between nearest-neighbor spins. In the second stage, we assume that the system is in the presence of the all introduced magnetic fields consisting of a finite inhomogenity. The magnetization, susceptibility and the specific heat can be straightforwardly calculated from  the Gibbs free energy $f$  according to the basic thermodynamic relations
\begin{equation}
\begin{array}{lcl}
 {M}=-\big(\frac{\partial f}{\partial B}\big), \quad {\chi}=\frac{\partial M}{\partial B},\quad {C}=-k_BT\big(\frac{\partial^2 f}{\partial T^2}\big),
\end{array}
\end{equation}
The non-conserved magnetization can be directly interpreted using an unusual behavior of the magnetic susceptibility at low temperature.
Figure \ref{fig:MXB1} displays exact results for the magnetization and magnetic susceptibility as a function of the longitudinal magnetic field $B_z/J_z$ for various fixed values of the transversal field $B_x/J_z$ at low temperature, where the Heisenberg coupling constants have been conventionally taken as $J_{x}=8J_z$ and $J_{y}=10J_z$ (one may consider different values for these parameters in order to more diversity of investigations). In this figure we consider the model under homogeneous magnetic fields (inhomogeneous parameters $b_z/J_z$, $b_x/J_z$, and $\lambda/J_z$ are equal to zero) with finite lengths of $N=6$ and $N=10$, separately. Panels \ref{fig:MXB1}(a) and  \ref{fig:MXB1}(c) show the magnetization and susceptibility of the spin chain with length $N=6$. At low temperature, weak transverse magnetic field $B_x/J_z$ and low transverse staggered field with $\theta=\frac{\pi}{30}$ (black dot-line),  there is a plateau at zero magnetization $M / M_s=0$, as well as, two intermediate plateaus at $M / M_s=\frac{1}{3}$ and $M / M_s=\frac{2}{3}$. With the increase of $B_x/J_z$, magnetization plateaus gradually convert to their counterpart quasi-plateaus. Although, the transformation from plateau to quasi-plateau will speed up upon increasing angle $\theta$ (the inset of Fig.  \ref{fig:MXB1}(a)). The transverse field $B_x/J_z$ and angle $\theta$ increment leads to delay in reaching saturation magnetization (see blue solid-lines).

In Fig. \ref{fig:MXB1}(c) the magnetic susceptibility for the same set of parameters is shown. The susceptibility behavior evidences the non-plateau nature of the magnetization within the same eigenstates of the model. Interestingly, for the strong field $B_x/J_z$ one can see that the susceptibility monotonically decreases upon increasing the field $B_z/J_z$ (blue solid-line). With further increase of the field $B_z/J_z$, the susceptibility has a non-monotone behavior with the maximums in those intervals of the longitudinal magnetic field at which quasi-plateaus arise in the magnetization curve.  This difference in behavior of the susceptibility for various fixed values of the transverse field $B_x/J_z$ at low temperature indicates that the system undergoes several phase transitions by increasing longitudinal field $B_z/J_z$.

Panels \ref{fig:MXB1}(b) and \ref{fig:MXB1}(d) illustrate the magnetization and magnetic susceptibility for the chain of length $N=10$ under the same conditions as $N=6$. Here, in addition to the zero-magnetization plateau, there are four intermediate plateaus at: $M / M_s=\frac{1}{5}$, $M / M_s=\frac{2}{5}$, $M / M_s=\frac{3}{5}$, $M / M_s=\frac{4}{5}$, then the magnetization reaches its saturation in strong magnetic fields. As a result, while  the number of magnetization plateaus increases with increase of chain size, the effect of transverse magnetic field $B_x$ and angle $\theta$ is more sensible in this case. Namely, the transformation from plateau to quasi-plateau occurs for the lower amount of applied field $B_x/J_z$. To clarify this point, if one compares red dash-dotted lines drawn in panels \ref{fig:MXB1}(a) and  \ref{fig:MXB1}(b) together, he finds that for the case $N=10$ quasi-plateaus appear for lower amounts of the transverse field compared with that of for case $N=6$.  As before, by increasing 
$\theta$ quasi-plateaus gradually disappear.

The magnetic susceptibility of the chain with $N=10$ as function of the longitudinal field $B_z/J_z$ for several fixed values of the transverse field is depicted in panel  \ref{fig:MXB1}(d). When the transverse magnetic field $B_x/J_z$ increases, the height of peaks of the susceptibility corresponding to the magnetization jumping between plateaus, decrease. As an important result, when the magnetization quasi-plateaus gradually appear by increasing the transverse field $B_x/J_z$, accordingly, special peaks of susceptibility will arise. We would like to draw your attention to another interesting effect of the transverse field increment on the susceptibility behavior, i.e., when the transverse field increases, the zero-field susceptibility has non-monotone behavior for both cases $N=6$ and $N=10$. The anomalous magnetic susceptibility behavior at very weak magnetic field 
$B_z/J_z$  is a remarkable evidence of existing magnetization quasi-plateau in the magnetization curve at low temperatures. With increase of the transverse staggered field coefficient $\theta$, the zero-field susceptibility gets further than other peaks in both cases $N=6$ and $N=10$. For the strong magnetic field region $B_z>10J_z$, there is a steep decrease in the susceptibility curve for all considered fixed values of the transverse field $B_x/J_z$, which denotes the magnetization goes to its saturation value.

\begin{figure}
\begin{center}
\resizebox{0.4\textwidth}{!}{%
\includegraphics{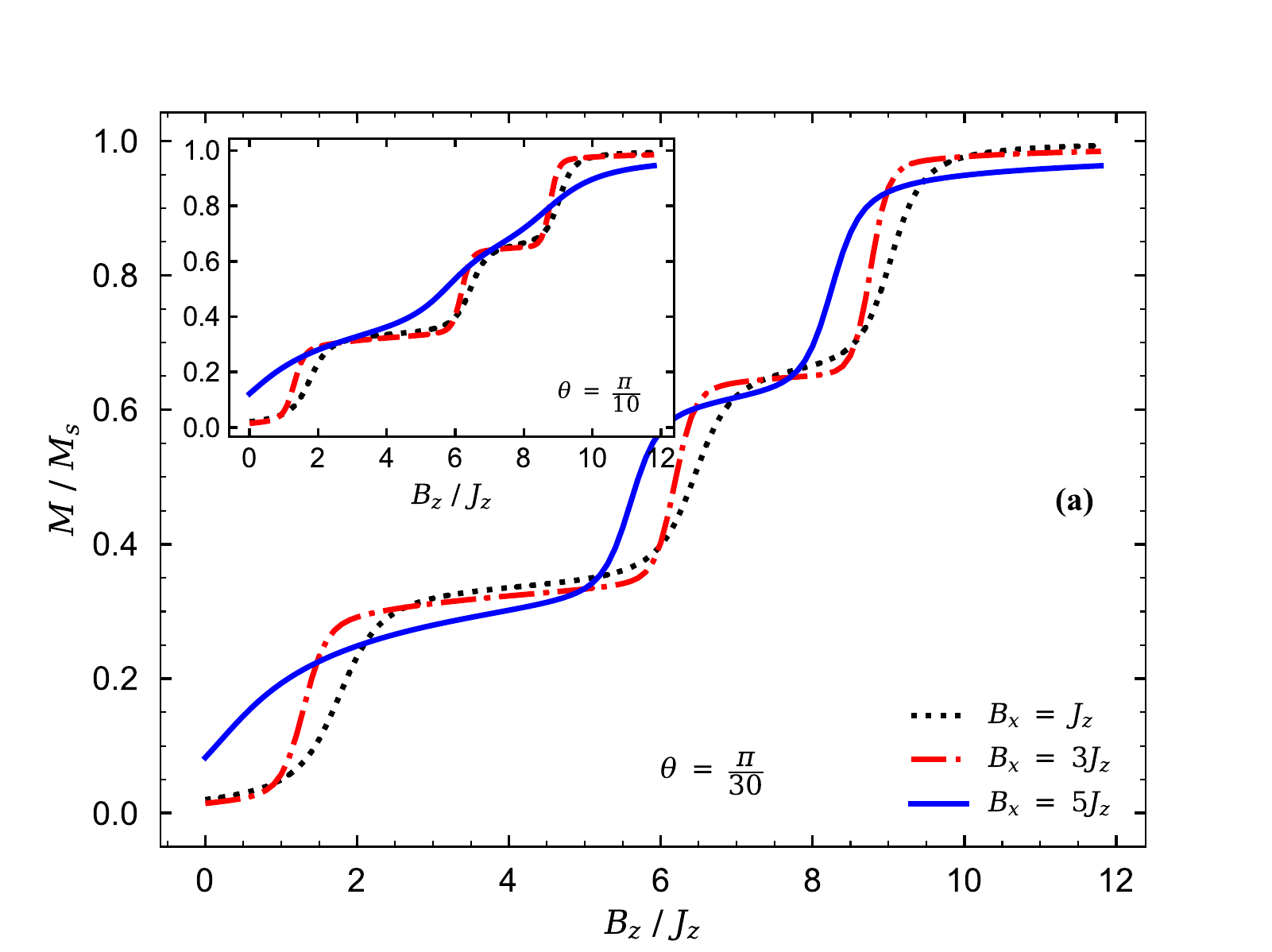}
}
\resizebox{0.4\textwidth}{!}{%
\includegraphics{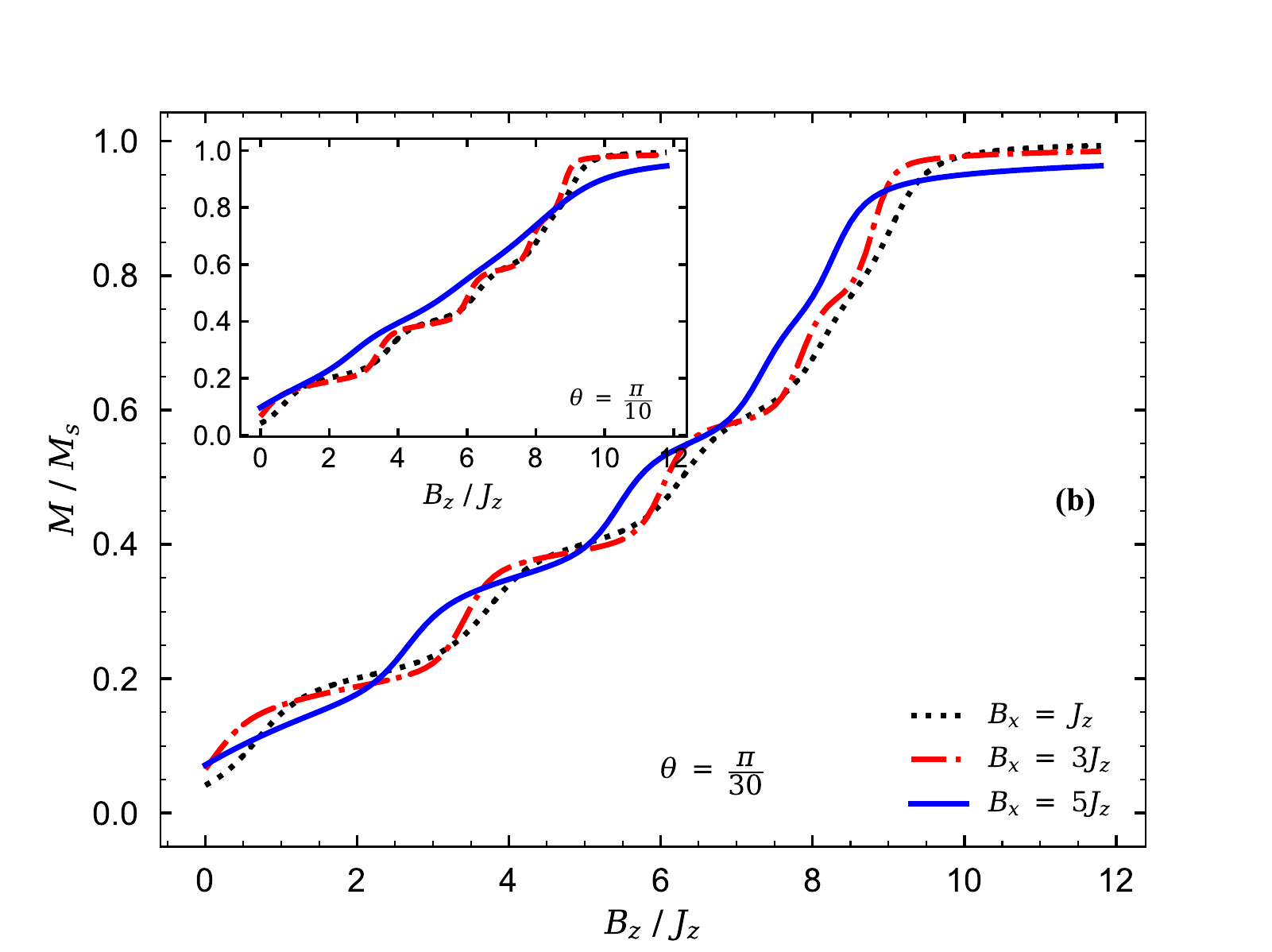}
}
\resizebox{0.4\textwidth}{!}{%
\includegraphics{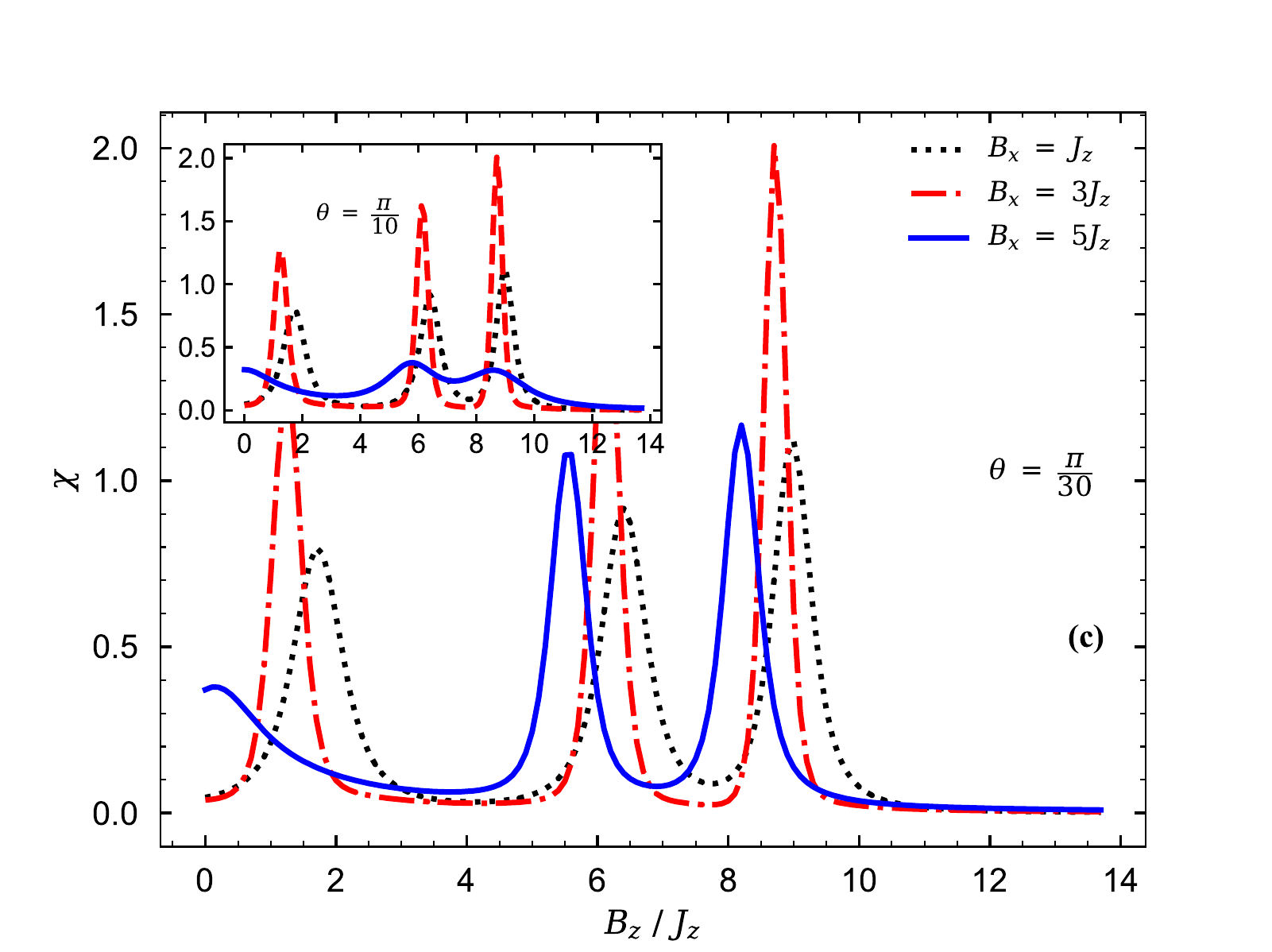}
}
\resizebox{0.4\textwidth}{!}{%
\includegraphics{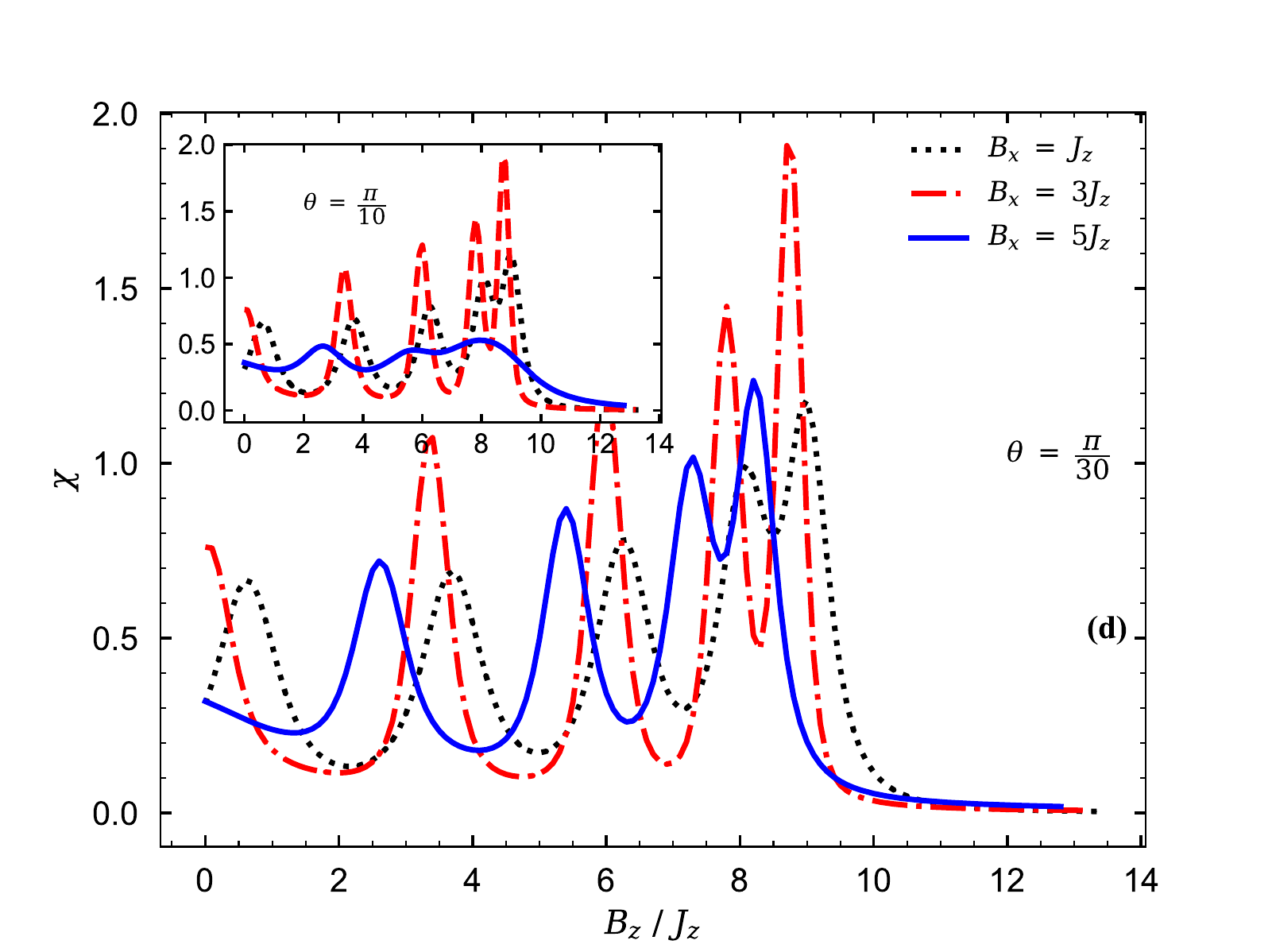}
}

\caption{Magnetization and magnetic susceptibility as functions of the longitudinal magnetic field $B_z/J_z$ for several fixed values of the transverse field $B_x/J_z$ at low temperature $(T=0.1J_z)$ and finite angle $\theta=\frac{\pi}{30}$. Inhomogeneous property is considered for all applied magnetic fields such that;  $b_z=0.6J_z$, $b_x=0.3J_z$, and $\lambda=J_z$. Coupling constants have been set as Fig. \ref{fig:MXB1}. Panels (a) and (c) are associated to the chain with finite length $N=6$; panels (b) and (d) correspond to that of the length $N=10$. Insets show the corresponding magnetization and magnetic susceptibility curves for higher transverse staggered field $B_x/J_z$ by setting $\theta=\frac{\pi}{10}$. }
\label{fig:MXB2}
\end{center}
\end{figure}

In order to accomplish with our discussion concerning to  finite-size effects of inhomogeneity property on the thermodynamic properties,
we study the behavior of the magnetization and magnetic susceptibility when the system is in the presence of inhomogeneous magnetic fields at low temperature. We have plotted in Fig. \ref{fig:MXB2} the magnetization and magnetic susceptibility of the model with the same conditions as figure \ref{fig:MXB1}, but under inhomogeneous magnetic fields (here, inhomogeneous parameters are taken as non-zero fixed values $b_z=0.6J_z$, $b_x=0.3J_z$, and $\lambda=J_z$). Panels \ref{fig:MXB2}(a) and  \ref{fig:MXB2}(c) display the magnetization and susceptibility with the finite length $N=6$ under inhomogeneous longitudinal, transverse, and transverse staggered magnetic fields. Panels \ref{fig:MXB2}(b) and  \ref{fig:MXB2}(d) are related to the chain of length $N=10$.
 In this situation  for both cases $N=6$ and $N=10$, all plateaus have been shifted toward higher values of the magnetization. Hence, we can see that inhomogeneity dramatically affects on the height and position of the low-temperature peaks in susceptibility. When the transverse magnetic field increases, firstly height of the peaks increases, then with further increase of the field $B_x/J_z$ gradually decreases. Moreover, under inhomogeneous magnetic fields, the susceptibility does not vanish even at zero longitudinal field $B_z=0$.
 Consequently, by imposing weak inhomogeneity property into the all magnetic fields, width of the magnetization plateaus decreases, and there is no zero magnetization plateau as well as zero-field susceptibility for the model under consideration with arbitrary length at low temperature. 
 
 By altering transverse staggered field intensity ($\theta=\frac{\pi}{10}$), one can see less variation in the shape of susceptibility for weak longitudinal field $B_z<2J_z$ compared with the case when the system is putted in the presence of homogeneous magnetic fields (see insets of figure \ref{fig:MXB2}).
It is quite evidence that under inhomogeneity, variations of both transverse field $B_x/J_z$ and transverse staggered field $B_x^{stagg}/J_z$ qualitatively affect the quasi-plateaus arisen in the magnetization curves more than when the system is in the vicinity of homogeneous magnetic fields, revealing that the magnetization curves including quasi-plateaus are more monotone than without inhomogeneity. 
\begin{figure}
\begin{center}
\resizebox{0.45\textwidth}{!}{%
\includegraphics{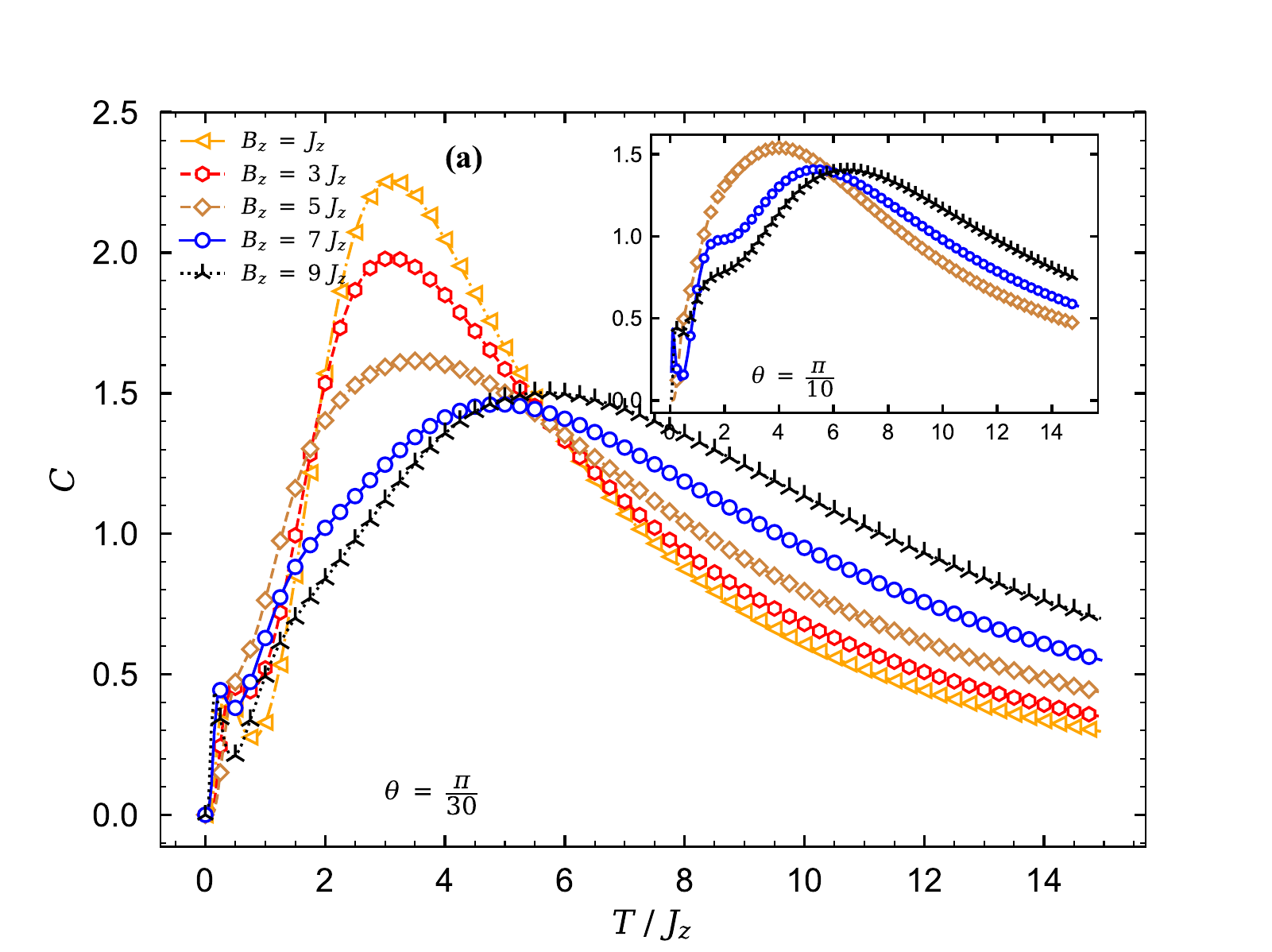}
}
\resizebox{0.45\textwidth}{!}{%
\includegraphics{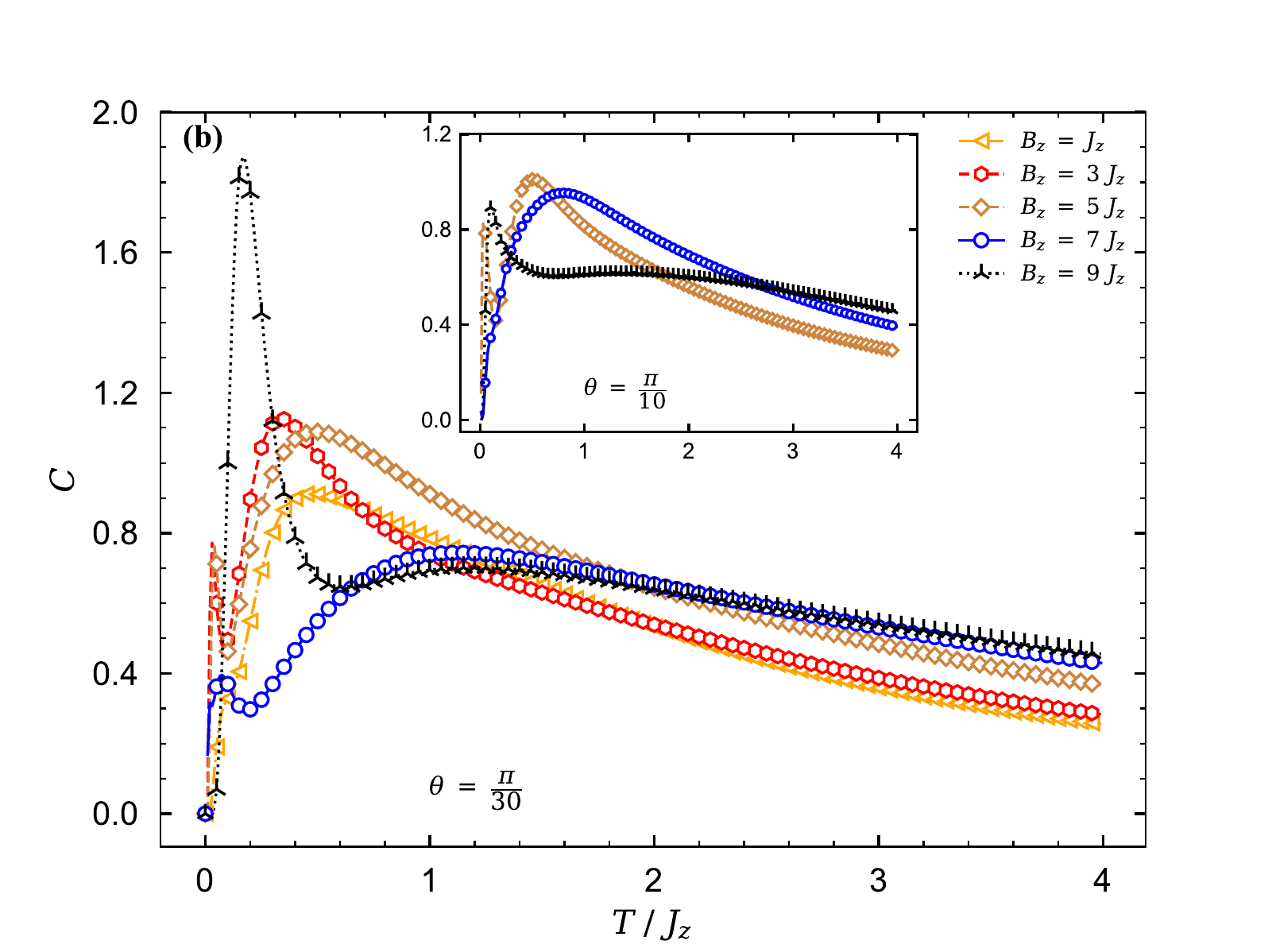}
}
\caption{Temperature dependences of the specific heat of the 1-D XXZ spin chain under various fixed values of the longitudinal magnetic field 
$B_z/J_z$. Other external magnetic fields and parameters are taken as $B_x=J_z$, $\theta=\frac{\pi}{30}$, $J_{x}=8J_z$ and $J_y=10J_z$. All applied magnetic fields are also considered as homogeneous fields such that: $b_z=0$, $b_x=0$, and $\lambda=0$. (a)  The specific heat of the chain with finite length $N=6$, and (b) $N=10$. Insets show the specific heat of the model in the presence of higher transverse staggered field as $\theta=\frac{\pi}{10}$. }
\label{fig:CT1}
\end{center}
\end{figure}

\begin{figure}
\begin{center}
\resizebox{0.45\textwidth}{!}{%
\includegraphics{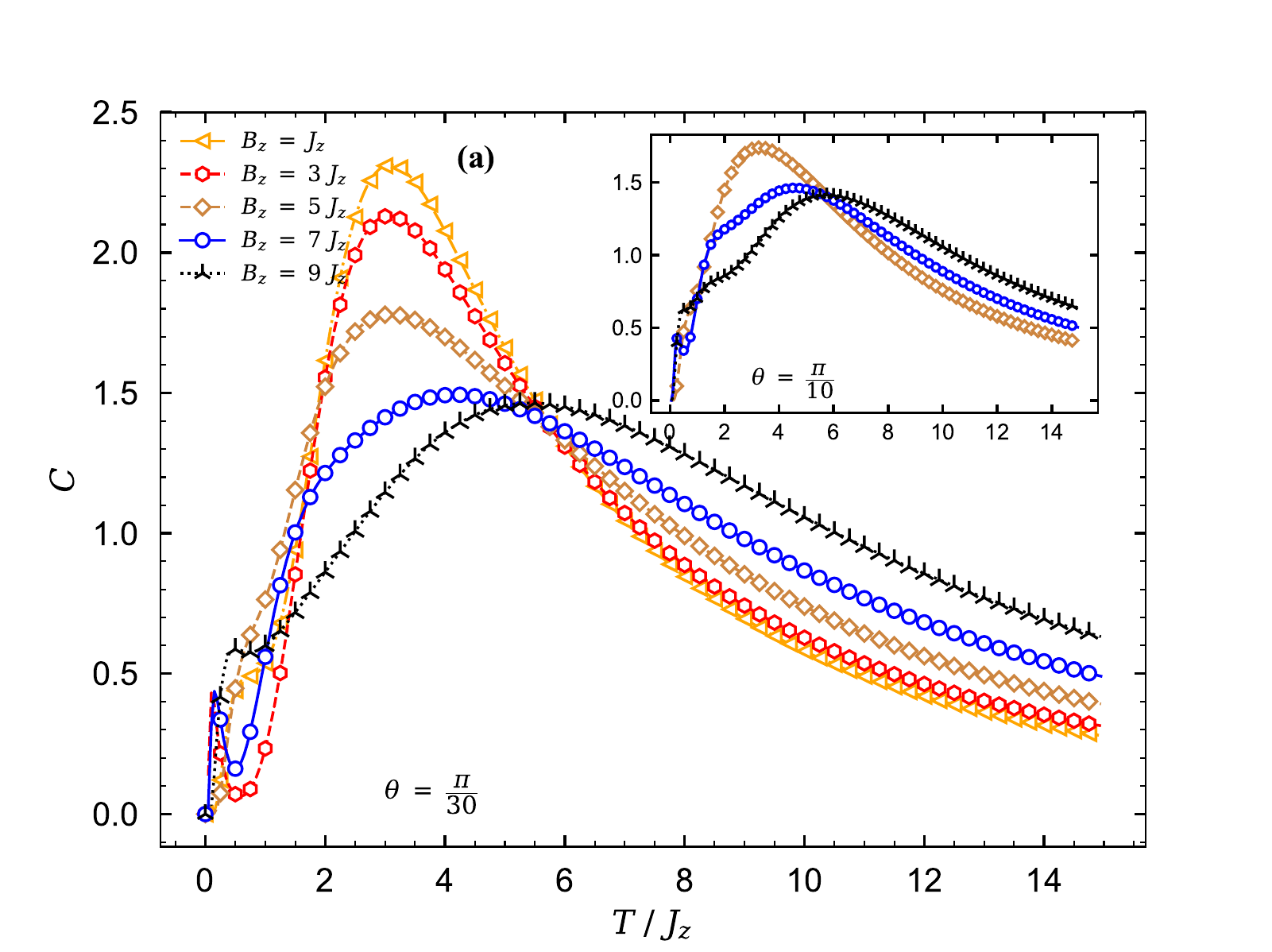}
}
\resizebox{0.45\textwidth}{!}{%
\includegraphics{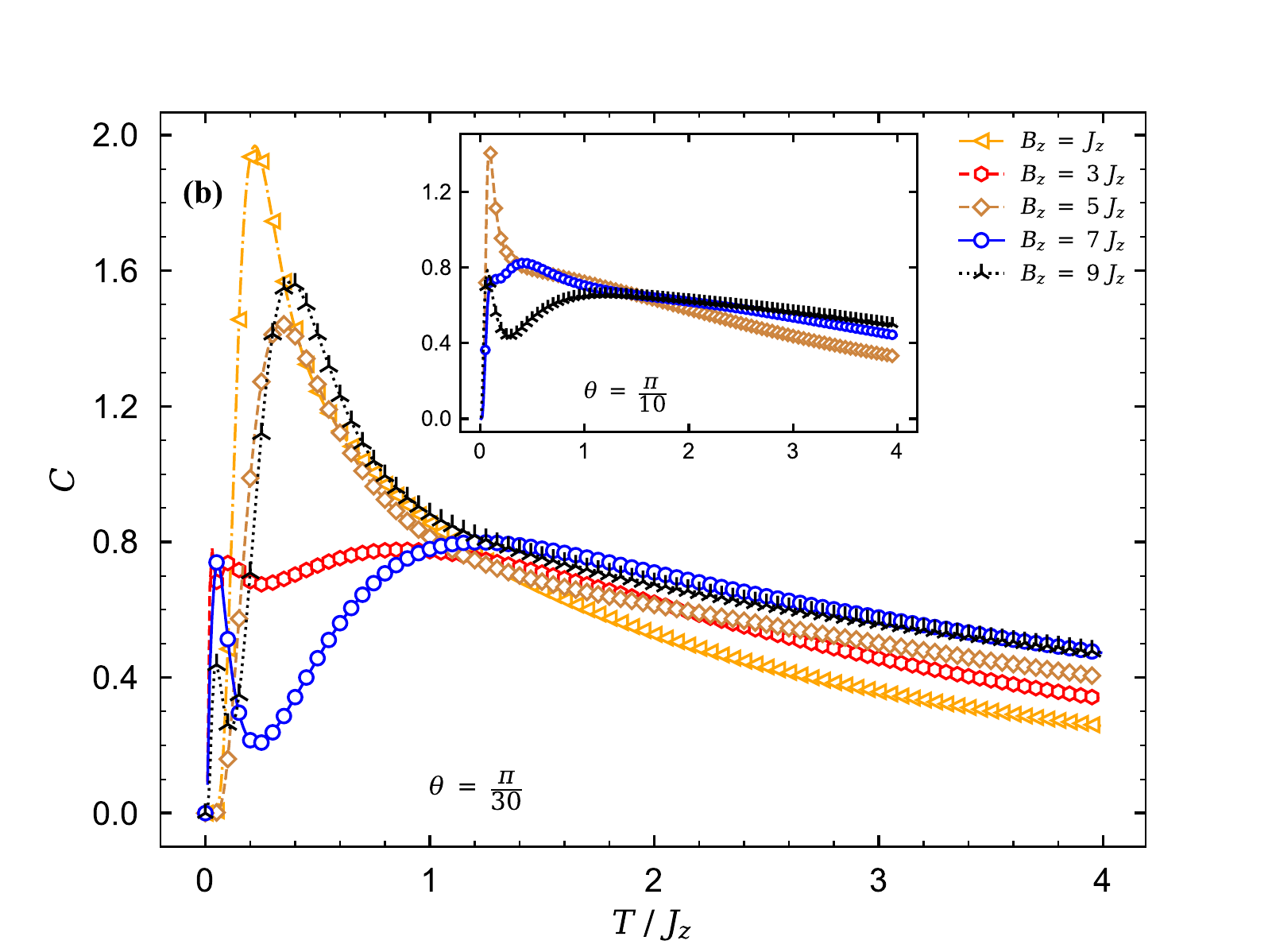}
}
\caption{Temperature dependences of the specific heat of the 1-D XYZ spin chain under various fixed values of the longitudinal magnetic field 
$B_z/J_z$. Other external magnetic fields and parameters are as in Fig. \ref{fig:CT1}. Here, all applied magnetic fields have been considered as inhomogeneous fields such that: $b_z=0.6J_z$, $b_x=0.3J_z$, and $\lambda=J_z$. (a)  The specific heat as function of the temperature for the chain with finite length $N=6$, and (b) $N=10$. Insets show the specific heat of the model versus temperature in the presence of higher transverse staggered field as $\theta=\frac{\pi}{10}$.}
\label{fig:CT2}
\end{center}
\end{figure}

Finally, we investigate the temperature dependences of the specific heat under both homogeneous and inhomogeneous external magnetic fields. 
The corresponding plots of the specific heat as function of the temperature for several fixed values of the longitudinal magnetic field are presented in Figs. \ref{fig:CT1} and \ref{fig:CT2}. When the chain of length $N=6$  (figure \ref{fig:CT1}(a)) is putted in the presence of homogeneous magnetic fields, it is seen that the specific heat exhibits a double-peak temperature dependence for weak longitudinal field $B_z\leq J_z$ at low temperature $T=0.1J_z$, where other parameters utilized in the Hamiltonian are taken as $B_x=J_z$, $b_z=0$, $b_x=0$, $\lambda=0$, and $\theta=\frac{\pi}{30}$.%single Schottky-type maximum
The height of double-peak monotonically decreases with increasing the field $B_z/J_z$. With further increase of the longitudinal field $B_z/J_z$ double-peak merge together and create a broad single Schottky-type maximum with smaller height. In the high longitudinal magnetic fields ($B_z>5J_z$), one observes that Schottky-type peak convert to a double-peak (blue and black marked lines of Fig. \ref{fig:CT1}(a)).
Consequently, the shape of specific heat maxima alternatively change upon increasing the field $B_z/J_z$. When the transverse staggered field increases ($\theta=\frac{\pi}{10}$), the longitudinal field dependences of the specific heat are explicitly impressed (the inset of Fig. \ref{fig:CT1}(a)). In other words, varying the angle $\theta$ results in arising third peak in the strong longitudinal fields (blue and black marked lines in the inset of Fig. \ref{fig:CT1}(a)). 

For the chain with more sites ($N=10$), there is a double-peak in the specific heat curve for the range $B_z\leq 7J_z$). In this case, the specific heat maxima have an alternating behavior upon increasing $B_z/J_z$. Ultimately, we see that two maxima merge together and make a sharp and narrow Schottky-type maximum in the strong longitudinal  field $B_z/J_z$ (black marked line in Fig. \ref{fig:CT1}(b)). Increase of the angle $\theta$ also alters the shape, position and height of the peaks (the inset of Fig. \ref{fig:CT1}(b)). We note that in this case the behavior of specific heat maxima (from change in heights and positions point of view) versus altering  longitudinal  field $B_z/J_z$ is more regular rather than the case $N=6$. The relationship between Schottky peak and the double-peak can be plausibly identified in terms of the alternations of the inhomogeneity parameters.

Let us now examine the specific heat for the case when the system is in the presence of external inhomogeneous magnetic fields. As shown in Fig.  \ref{fig:CT2}(a), by imposing inhomogeneity property into the applied magnetic fields as $b_z=0.6J_z$, $b_x=0.3J_z$, and $\lambda=J_z$, where $B_x=J_z$ and $\theta=\frac{\pi}{30}$, one can see a double-peak for the range $B_z\leq 5J_z$. As we noted for Fig. \ref{fig:CT1}(a), by increasing the field $B_z/J_z$, double-peak gradually merge together and finally make a single Schottky-type maximum in the range
$3J_z<B_z<7J_z$. With further increase of  $B_z/J_z$, the double-peak appears again in the specific heat curve.
Another important point affecting the maxima of the specific heat is the altering the angle $\theta$. For higher values of  $\theta$, we see that the specific heat displays a double-peak for the strong magnetic field $B_z/J_z$ and fixed $B_x=J_z$, which merge together by decreasing the longitudinal field.  For higher transverse staggered field ($\theta=\frac{\pi}{10}$), we witness an anomalous triple-peak temperature dependence in the presence of strong longitudinal magnetic field $B_z\geq 7J_z$.

When the number of sites in the chain increases (Fig. \ref{fig:CT2}(b)), there is a sharp Schottky-type maximum for weak longitudinal magnetic field.
By increasing the magnetic field $B_z/J_z$, a double-peak will appear and remains even for strong longitudinal magnetic fields. When the transverse staggered field coefficient $\theta$ increases, the shape of Schottky-type maximum remarkably changes, namely, it gets more sharper and narrow with lower temperature position. When the strength of longitudinal magnetic field increases a double-peak arises for values $B_z\leq 7J_z$ (the inset of Fig. \ref{fig:CT2}(b)).

\section{Conclusions}\label{conclusions}
The present work deals with the study of magnetization, magnetic susceptibility and the specific heat of the exactly solvable 1-D spin-1/2 XYZ chain under different external magnetic fields including longitudinal, transverse, and transverse staggered. Two small number of particles have been considered for the chain under periodic boundary conditions due to better understanding the finite-size effects of inhomogeneity on the spin models. The thermodynamic parameters of the spin system have rigorously been investigated under two different circumstances: Firstly, for the case when the system is in the presence of homogeneous magnetic fields; Secondly, for the case when all applied magnetic fields have a finite inhomogeneiny property. To consider suitable inhomogeneity properties in the applied magnetic fields, we have implemented inhomogeneity coefficients correspond to the three kinds of applied magnetic fields consisting of longitudinal, transverse, and transverse staggered magnetic fields. As a matter of fact, we have assumed that the strength of the induced magnetic fields is different for the odd and even sites of the chain. Since  the magnetization operator does not commute with the Hamiltonian some unusual phenomena have been observed.

The low temperature examinations of the magnetization and magnetic susceptibility for the XYZ model under homogeneous magnetic fields revealed that the magnetization curve undergoes an interesting evolution such that upon increasing the transverse magnetic field, all plateaus convert to their counterpart quasi-plateaus. Moreover, by increasing the staggered field coefficient $\theta$, quasi-plateaus gradually disappear, where the magnetization has a smooth curve versus longitudinal magnetic field for the high values of the transverse field and larger
 $\theta$. We have observed that the susceptibility curve has also intriguing behavior with respect to the longitudinal magnetic field, when the strength of the transverse and transverse staggered fields change. In a good agreement with the jumps between magnetization plateaus, susceptibility curve has maxima whose shapes and positions are strongly dependent on the strength of all applied fields in the Hamiltonian. The non-monotone behavior of the susceptibility at higher values of transverse field indicates existence of the quasi-plateaus in the magnetization at low temperature. We also found a zero longitudinal field susceptibility upon increasing the transverse field. When the inhomogeneity property was imposed  into the magnetic fields, low temperature behavior of the magnetization and magnetic susceptibility versus longitudinal field remarkably varied for low amounts of the transverse magnetic field. As a main result, here we have seen a zero longitudinal field susceptibility for both cases $N=6$ and $N=10$ even under the weak transverse magnetic field.

Finally, we have examined the specific heat of the finite-size model as function of the temperature for various fixed values of the longitudinal magnetic field.
We have concluded that, when the system with $N=6$ particles is putted in the presence of homogeneous magnetic fields, there is a strongly transverse field dependent double-peak which gradually tends to a Schottky-type maximum  upon increasing  the transverse field.
Amazingly, for the strong longitudinal magnetic field, a triple-peak have been appeared in the specific heat curve when the strength of the transverse staggered field increased, while for the case $N=10$ we have not observed triple-peak even at high  longitudinal and high transverse staggered fields. Our calculations and simulations prove that changes in the specific heat behavior are in an excellent coincidence with the magnetization steps and jumps, accompanying with the magnetic ground-state phase transitions.

 for the case when the system is considered under inhomogeneous magnetic fields, the shape and the positions of the specific heat maxima have been remarkably changed specifically for chain of $N=10$ particles. Ultimately, we understood that for the finite length chains the altering staggered field has substantial influences on the behavior of the magnetization process, magnetic susceptibility and specific heat under the both different circumstances described above. Our exact results and straight expressions demonstrated in this work are fundamentally applicable for investigating small spin clusters and single molecular magnets with the same size in the presence of different kinds of magnetic fields that are crucial not only in the theoretical condensed matter but also in the experimental activities.

\section*{Acknowledgments}
H. Arian Zad acknowledges the receipt of the grant from the ICTP Affiliated Center Program AF-04, and 
the CS MES RA in the frame of the research project No. SCS 18T-1C155.

\bibliography{}

\end{document}